\documentclass[12pt]{amsproc}
\numberwithin{equation}{section}
\def\half{\mbox{\scriptsize{${{1}\over{2}}$}}}

\def\thalf{\mbox{\scriptsize{${{3}\over{2}}$}}}
\def\gs{g_{\rm{s}}}
\def\ls{\ell_{\rm{s}}}
\def\enh{enhan\c con}

\def\nef{$n{=}4$}
\def\net{$n{=}2$}
\def\nets{$n{=}2*$}
\def\detn{$d{=}10$}
\def\bP{\mathbf{P}}
\def\bR{\mathbf{R}}
\def\vtpt{$\vartheta=\pi/2$\,}
\def\vpz{$\varphi=0$\,}
\newcommand{\ul}[1]{\underline{#1}} 
\begin{document}

\begin{flushright}
hep-th/0106148
\end{flushright}

\vbox{\vskip-2.0truein}

\title{More on Singularity Resolution}

\author{Amanda W. Peet}

\address{Author address: University of Toronto \\ Department of
Physics \\ 60 St. George St, Toronto, Ontario, Canada M5S 1A7.}

\begin{abstract}
String theoretic resolution of classical spacetime singularities is
discussed.  Particular emphasis is on the use of brane probes, and the
connection [1] of the \enh\ phenomenon [2] to the \nets\ Pilch-Warner
flow spacetime [3].  Some comments and details on the singularity of
the PW spacetime are added.  For the proceedings of Strings 2001,
Mumbai, India.
\end{abstract}

\maketitle
\section{Introduction}

Black holes have long been sources of puzzles in theoretical physics.
When treated semiclassically, black holes have two problematic
aspects: their event horizons give rise to information loss, and their
curvature singularities point to breakdown of the theory.  A quantum
theory of gravity is clearly needed to address both problems.

Classical spacetime singularities provide a valuable testing ground
for a quantum gravity theory \cite{garyrob}.  Singularities must be
dealt in one of two ways: (a) {\em resolution} via some short-distance
degrees of freedom not encoded in classical gravity, or (b) {\em
prevention}.  The existence of possibility (b) is important and oft
forgotten.  The essential physics behind it is that some singular
classical geometries are so sick that the quantum theory should not
allow them ab initio - or allow them to form starting from physical
initial conditions.
%
A simple example of a sick spacetime is Schwarzschild with mass $M<0$.
A resolution, no matter how innocuous it might look locally, would
violate stability of the vacuum.  Supersymmetry forbids it: GR can be
embedded into $n{=}1$ supergravity and a Bogomolnyi bound $M\geq 0$
derived \cite{witten1}.

Here, quantum gravity means string theory, the low-energy limit of
which is supergravity.  Construction of solutions of the highly
nonlinear equations of motion of supergravity, such as black holes and
branes, is typically difficult.  Any given search can be aided by a
no-hair theorem.  This guarantees uniqueness once the conserved
quantum numbers associated to the spacetime are fixed, provided that
cosmic censorship is not violated.  If a singularity is naked,
however, there is {\em no} guarantee that a spacetime solving the
equations of motion is the correct one.

In supergravity, all types of spacetime singularities are encountered:
cosmically censored and naked; spacelike, null, and timelike.  It is
generally hard to learn the quantum resolution mechanism for any given
classical singularity.  One reason is that a spacetime which appears
singular in a lower dimension may in fact become nonsingular when
lifted to higher dimension \cite{ght}.  For this reason it is best to
analyse singularities in {\em ten} dimensions.
Another word of warning is that whether or not a delta-function source
is necessary in the right-hand side of an equation of motion can be
coordinate-dependent.  
%

String theory knows what to do in extreme regimes, and so classical
spacetime singularities apparently occurring in spacetime geometries
arising from fundamental objects such as strings and D-branes must
have sensible resolutions.  It is therefore of considerable interest
to investigate geometries arising from combinations of fundamental
ingredients, and much work has been done which cannot possibly be
reviewed comprehensively here.  The \enh\ mechanism \cite{jpp} was one
notable example of singularity resolution; others included e.g.
\cite{polstr} via dielectric-brane expansion, \cite{klebstr} via a
supergravity resolution into fluxes and \cite{vaf} via geometric
transitions.  In the following, systems with \net\ supersymmetry will
be the focus of attention.

The \enh\ phenomenon \cite{jpp} arose in the \net\ supersymmetric
context of D-branes wrapped on K3 but it has other realisations.
Consider for example $N_5$ D5-branes wrapped on a K3 surface.  In this
spacetime, the K3 volume decreases with decreasing radius.  (This
behaviour is to be contrasted with the case where $N_1>N_5$ D1-branes
are added, giving rise to AdS$_3\times$S$^3\times$K3 spacetime and
fixed K3 volume in the interior.)  At the \enh\ radius, the K3 volume
goes to its self-dual value, and the tension of a wrapped D-probe
vanishes giving an enhanced gauge symmetry.  The D-probe cannot go
further in.  The $N_5$ source-branes have an \enh\ radius proportional
to $N_5$, and by induction it can be argued that they can never get
close enough to allow a naked singularity to form.  Instead, they live
on a spherical shell, by Gauss's law spacetime is flat inside, and the
would-be singularity is excised.  In the realisation and regime of
parameter space where there is a good gauge theory dual, the \enh\ can
be seen as a nonperturbative effect by analysing the Seiberg-Witten
curve.

Although the \enh\ mechanism of singularity resolution arises as an
essentially stringy phenomenon, in fact supergravity already knows
about it.  The precise details of the excision story from the point of
view of supergravity jump conditions were studied in \cite{jmpr}.
Since there is a moduli space for motion of a probe-brane in the
overall transverse directions, the shell of branes can also live at
any radius greater than the \enh\ radius.  The jump conditions for all
supergravity fields then prove to be self-consistently satisfied by
D-brane sources with tension exactly as appropriate to the running K3
volume.  Supergravity does not allow the branes to be packed closer
than the \enh\ radius: this would require unphysical negative tension.
Some details on non-extremal deformations of the \enh\ are also
contained in \cite{jmpr}.

\section{The  \nets\ Spacetime: Symmetries and Singularities}

In the context of the original AdS$_5$/CFT$_4$ correspondence, many
deformations were considered in which the CFT was perturbed by
relevant or marginal operators, and the supergravity solution was
changed correspondingly in the interior.  In \cite{pilwar} the
consistent truncation Ans{\"a}tz was used to produce a proposal for a
spacetime dual to the \nets\ flow, obtained by turning on vevs for the
field theory operators
$O_f = {\rm{Tr}}\left( \lambda^3\lambda^3+\lambda^4\lambda^4\right),
{\bar{O_f}}$
and
$O_b =  \sum_{j=1}^4 {\rm{Tr}}\left( X^i X^i\right)
    - 2\sum_{j=5}^6 {\rm{Tr}}\left( X^i X^i\right)$.
The resulting spacetime, which is abbreviated here as PW, has a small
number of integration constants.  On the other hand, the moduli space
for the \net\ gauge theory has $O(N)$ parameters, so the PW spacetime
should correspond to a small subspace of that moduli space.  The
parameters of the PW solution are labelled $\gamma$ and the
\nef-breaking parameter\footnote{$k$ is dimensionless; the
proportionality constant will be fixed in section 4.}  $k\propto mL$,
where $m$ is the mass parameter and $L$ the radius of curvature of the
asymptotic AdS$_5$ and S$^5$.  To uncover the link to \enh\ physics
the parameter $\gamma$ must be set to zero, the case of maximal
breaking of \nef\ in the PW solution.  All expressions in the
following will be written for $\gamma=0$ only.  Note that most
important feature of the PW spacetime as compared to previous `flow'
geometries was that the \detn\ dilaton-axion field varies with radius.
All analysis will be done in \detn\ to avoid confusion.

The \nets\ gauge theory has $SU(2)\times U(1)$ R-symmetry.  This is
easy to see by looking at the four Weyl fermions: $\lambda^{3,4}$ get
mass and an $SO(2)=U(1)$ mixes them, while $\lambda^{1,2}$ are
massless and mixed by an $SU(2)$.  The 6 transverse scalars $X^i$
transform as $(4\times 4)_A$, and two of them are invariant under the
R-symmetry.  Since gauge theory symmetries are spacetime isometries,
this condition on the $X's$ gives rise to a fixed-plane in the
spacetime where the radius of the transverse (squashed) sphere goes to
zero.  The R-symmetry does not act on the azimuthal angle $\varphi$;
the supergravity fields therefore can (and do) depend on it in
complicated fashion.  The spacetime possesses in addition an
accidental $U(1)'$ symmetry in Einstein frame.  This $U(1)'$ is a
combination of a $U(1)$ subgroup of the R-symmetry group and a $U(1)$
subgroup of the supergravity S-duality group $SL(2,\bR)$.

The relation of the PW radial coordinate to the familiar \nef\
isotropic coordinate $r$ is given via
\begin{equation}
\rho^6=c+\half\left(c^2-1\right)\ln\left({{c-1}\over{c+1}}\right) 
\,,\quad
c=\cosh\left(kL/r\right) \,.
\end{equation}
Accordingly, $c\in[1,\infty)$.

The PW spacetime has many fields turned on; the metric is in Einstein
frame
\begin{equation}\label{eq:pwein}
\begin{split}
&\! {{ds_E^2}\over{L^2}} =
{{(cX_1X_2)^{1/4}}\over{\rho^3}}
\left\{
   {{(k/L)^2\rho^6}\over{c^2-1}}dx_\parallel^2 
   +{{1}\over{\rho^6(c^2-1)^2}}dc^2 + 
   \left[ 
      {{1}\over{c}}d\vartheta^2 
   \right.
\right. 
\cr &\! 
\left.
   \left.
      +{{\sin^2\vartheta}\over{X_2}}d\varphi^2 
      +{{\rho^6\cos^2\vartheta}\over{4}} 
      \left(
         {{1}\over{X_1}}d\psi^2 + {{1}\over{cX_2}}d\alpha^2
         +{{2\cos\psi}\over{cX_2}}d\alpha d\beta 
      \right. 
   \right.
\right. 
\cr &\!
\left.
   \left.
      \left.
         +d\beta^2\bigl({{\sin^2\vartheta}\over{X_1}}
                        +{{\cos^2\vartheta}\over{cX_2}}\bigr) 
      \right)
   \right] 
\right\} 
\equiv 
\left[{{\sqrt{cX_1X_2}}\over{\rho^6}}\right]
^{1/2}\biggl\{ ds^2 \biggr\} \,.
\end{split}
\end{equation}
where
\begin{equation}
X_1=\cos^2\vartheta+c\rho^6\sin^2\vartheta \,,\quad
X_2=c\,\cos^2\vartheta+\rho^6\sin^2\vartheta \,.
\end{equation}
Meanwhile, the R-R self-dual 5-form field strength can be
written
\begin{equation}
F_{(5)}=f+*f \,,\quad f = 4dx^0\wedge dx^1\wedge dx^2\wedge dx^3
\wedge dw(c,\vartheta) \,,
\end{equation}
and both the R-R and NS-NS 3-form fields strengths are turned on in
directions transverse to the D3-branes only.  The dilaton-axion field
$\tau$ runs:
\begin{equation}\label{eq:taus}
\tau = {{\tau_0-{\bar{\tau_0}}B}\over{1-B}} \,, 
\quad{\rm where}\quad B = e^{2i\varphi} 
{{\sqrt{cX_1}-\sqrt{X_2}}\over{\sqrt{cX_1}+\sqrt{X_2}}} \,,
\end{equation}
and $\tau_0=i/\gs+\theta_s/(2\pi)$ is the asymptotic coupling.  

It is interesting to perform an analysis of the singularity of this
spacetime.  The above metric (\ref{eq:pwein}) is in Einstein frame.
For string probe physics the string frame metric is needed, and the
conversion $g^S_{\mu\nu}=e^{\Phi/2}g^E_{\mu\nu}$ involves only the
\detn\ dilaton.  Study of (\ref{eq:taus}) gives
\begin{equation}
e^\Phi = {{(cX_1{+}X_2)}\over{2\sqrt{cX_1X_2}}}-
\cos(2\varphi){{(cX_1{-}X_2)}\over{2\sqrt{cX_1X_2}}} \,.
\end{equation}
Note that this is normalised to unity far out at $c=1$.  Large string
loop corrections then occur when the dilaton is large, and it is not
hard to see that this occurs on \vtpt\ at large-$c$.

To compute curvatures in either frame it is simplest to separate off
the conformal factor and use the formula for
${\tilde{g}}_{\mu\nu}=\Omega^2 g_{\mu\nu}$
\begin{equation}\label{eq:romega}
{\tilde{R}}={{1}\over{\Omega^2}}\left[R-2(d-1)\nabla^2 \ln\Omega
- (d-1)(d-2)\left(\nabla \ln\Omega\right)^2 \right] \,.
\end{equation}
Now, the conformal factors entering the scalar curvature formula
(\ref{eq:romega}) are, for Einstein and string frame respectively,
\begin{equation}
\Omega_E^4 = {{\sqrt{cX_1X_2}}\over{\rho^6}} \,,\quad 
\Omega_S^4 = {{1}\over{2\rho^6}} \left[ \left(cX_1{+}X_2\right)
-\cos(2\varphi)\left(cX_1{-}X_2\right) \right] \,.
\end{equation}
The Ricci scalar curvature for the plain metric, the part inside the
$\{ \}$ of (\ref{eq:pwein}), can be easily found.  Expressions for
this and other ingredients are cumbersome but simplify somewhat in the
large-$c$ limit\footnote{In all large-$c$ scalings presented here
there appears in various contributions to the Ricci scalar a ratio of
(sometimes rational powers of) polynomials of angles, which cannot be
expanded here for lack of space.  These functions are denoted $f_i$
and have nothing to do with $f$ in the R-R five-form.}.  At large-$c$,
$R\sim c f_0(\vartheta)$ generically and $R\sim 30c^3$ on \vtpt.

In the large-$c$ limit, the Einstein conformal factor behaves as
$\Omega_E^4\sim c^2 f_1(\theta)$ generically and $\Omega_E^4\sim c$ on
\vtpt.
For the string frame at large-$c$, 
$\Omega_S^4\sim c^2 f_2(\vartheta,\varphi)$ generically and
$\Omega_S^4\sim 1$ on (\vtpt, \vpz).
Thus the $1/\Omega^2$ suppression in the curvature in either Einstein
or string frame, at large-$c$, is less powerful on the above
non-generic loci, and this will result in more singular behaviour
there.
The other contributions to the Ricci scalar involve derivatives of the
conformal factor.  For the Einstein frame, the d'Alembertian term
scales as
$c f_3(\vartheta)$ generically and $\thalf c^3$ on \vtpt;
the gradient-squared term scales as 
$c f_4(\vartheta)$ for all $\vartheta$.  
In string frame, the d'Alembertian term scales as
$c f_5(\vartheta,\varphi)$ generically and 
$3 c^3$ on (\vtpt, \vpz);
the gradient-squared term scales as 
$c f_6(\vartheta,\varphi)$ generically and vanishes (!) on (\vtpt,
\vpz).

Overall, for the large-$c$ scalings, the Einstein frame results are
\begin{equation}
R_E \rightarrow 
\left\{ \begin{split}
f_7(\vartheta) \,, \vartheta\not=\pi/2 \cr
3 c^{5/2} \,, \vartheta=\pi/2 \cr
\end{split} \right.
\end{equation}
while in string frame the results are
\begin{equation}
R_S \rightarrow
\left\{ \begin{split}
f_8(\vartheta,\varphi)  \,, \vartheta\not=\pi/2,\varphi\not=0 \cr
c^2 f_9(\varphi)\,, \vartheta    =\pi/2,\varphi\not=0 \cr
-24 c^3\,, \vartheta    =\pi/2,\varphi    =0 \cr
\end{split} \right.
\end{equation}
Therefore, curvature invariants are finite generically but blow up as
$c\rightarrow\infty$ on the locus \vtpt, signalling the need for
$\alpha'$ corrections.

\section{Brane Probes and Supergravity}

Generically, D-brane probes feel different physics than gravitons or
other excitations of supergravity fields, because they couple
differently.  In the case of the original \enh\ phenomenon, the
physics became clearest upon probing with a brane identical to the
source-branes.  In particular, it became manifest that the
source-branes could not get close enough together to allow formation
of a naked classical singularity.

The same ideas can now be used on the PW spacetime, in order to find
the connection to the original \enh\ phenomenon, and to uncover the
constituent structure of the source-branes giving rise to the
spacetime.  The natural Ans{\"a}tz is to take the constituents to be
D3-branes with {\em no} higher-brane multipoles, with a distribution
to be calculated.  Again, to avoid confusion, the probe analysis must
be done in \detn.  Details not shown explicitly here may be found in
the original paper \cite{bpp} (related work is in \cite{evjohpet}).

The D$p$-brane probe action is, in general,
\begin{equation}
\begin{split}
S_{\rm probe} = & -{{\mu_p}\over{\gs}} \int d^{p+1}\xi \ e^{-\Phi}\ 
\sqrt{\det\left(\bP\left[g+B\right]_{ab}+2\pi\ls^2F_{ab}\right)}\cr
& +{\mu_p}\int \bP\exp\left(2\pi\ls^2F_{(2)}+B_{(2)}\right)\wedge
\oplus_n C_{(n)} \,,
\end{split}
\end{equation}
where $\bP$ denotes pullback of bulk fields to the brane worldvolume.
It is simplest to work in static gauge.  Now, probe physics in the PW
background is simpler than it first appears because the $\bP B_{(2)},
\bP C_{(2)}, F_{(2)}$ terms vanish, and the cross-terms in $\bP
F_{(5)}$ are absent.

The potential energy function vanishes on two loci.  The first is the
equator of the $S^5$, i.e. \vtpt.  The second is the special radius
$\rho^6=0$; the coordinate $c\rightarrow\infty$ there.  Two coordinate
patches are needed for the fixed-plane required by the R-symmetry: the
$(\rho,\vartheta)$ plane and $(\vartheta,\varphi)$ with identification
$\vartheta\simeq\pi-\vartheta$.  This gives a two-dimensional moduli
space, in accordance with expectations from the \net\ gauge theory.

The physics becomes clearer when a change is made to a coordinate
system appropriate to the \net\ structure.  This involves writing the
kinetic energy on the moduli space as
\begin{equation}
T(Y) = \half \tau_3 e^{-\Phi} v^Y v^{\bar{Y}} \,,
\end{equation}
for some complex field $Y$.  The transformation is
\begin{equation}
Y={{kL}\over{2}} \left(z+{{1}\over{z}} \right) \,, \quad {\rm{where\
}} z=e^{-i\varphi}\sqrt{(c+1)/(c-1)} \,.
\end{equation}
In these coordinates, the dilaton-axion field becomes
\begin{equation}\label{eq:Ytau}
\tau(Y) = {{i}\over{\gs}} \sqrt{{Y^2}\over{Y^2-k^2L^2}} + 
{{\theta_s}\over{2\pi}} \,.
\end{equation}
This is, as expected, a holomorphic function of $Y$.  
The probe's kinetic energy vanishes on a particular locus, signalling
the presence of an \enh.  This locus is a {\em line segment}, going
from $Y=-kL$ to $Y=+kL$, along the branch cut\footnote{Notice that in
(\ref{eq:Ytau}) it may appear at casual glance that the dilaton is
zero on the \enh\ locus.  In fact, the $Y$-dependent part of the
expression for $\tau(Y)$ has a real part, and once this is separated
off it is easy to see that the imaginary piece does in fact exhibit
the expected dilaton-blowup behaviour.}.  In coordinates appropriate
to the \net\ structure, then, the \enh\ is a line segment and not a
circle.

If $\gamma$ had been kept as a parameter here, it would be easy to
show that as $\gamma\rightarrow-\infty$, the line segment unsquashes
and the brane distribution goes over to a disc, as appropriate to the
\nef\ Coulomb branch problem.  Note also that this line segment is not
the \nets\ limit of the $n{=}1*$ story of \cite{polstr}; a different
perturbation is considered here.

\section{Gauge Theory Connection and Brane Distribution}

One advantage of working in coordinates appropriate to the \net\
structure is that it becomes easy to extract the distribution of
branes using
\begin{equation}
\tau_{\rm SUGRA}=\tau_{\rm SYM} \,.
\end{equation}
In computing the coupling function $\tau_{\rm SYM}$ for a Coulomb
branch configuration, it is best to begin by recalling from the
Seiberg-Witten story that quantum mechanically $SU(N)$ is always
broken down to $U(1)^{N-1}$.  The vevs in energy units, $a\equiv
Y/(2\pi\ls^2)$, can then be parametrised as
diag$(\left\{a_i\right\})$, where $\sum_i a_i=0$.  The next step is to
compute the prepotential $F$ for the $N-1$ abelian vector multiplets,
\begin{equation}
F=F_{\rm classical}+ F_{\rm pert} + F_{\rm nonpert} \,.
\end{equation}
The perturbative correction to the classical prepotential is 1-loop
exact, and is computed by integrating out the charged fields
(``W''-bosons).  The nonperturbative part is generated by instantons
and is difficult to calculate for large-$N$, but it is important only
when there are light BPS states, i.e. when the eigenvalue spacings are
smaller than order $1/N$.  Such effects are exponentially suppressed
in the supergravity regime where $N$ is large.  Nonperturbative
corrections turn on sharply at the \enh\ locus, where the spacing
between the eigenvalue representing the probe and those representing
the source-branes can in fact become close.

To find the perturbative part of the prepotential, place a probe at
distance $u$.  Then the vevs are changed to
diag$(u,\left\{a_i-u/N\right\})$, giving $F_{\rm pert}$ and thence
\begin{equation}
\tau_{\rm SYM}(u)= {{i}\over{\gs}} + {{\theta_s}\over{2\pi}}
+ {{i}\over{2\pi}} \sum_i \ln\left[ 
{{\left(u-a_i-u/N\right)^2}\over{\left(u-a_i-u/N\right)^2-m^2}}
\right] \,.
\end{equation}
The \nef\ breaking is small in the \nets\ solution of PW.  This means
that the logarithm can be Taylor expanded.  In the large-$N$ limit,
the sum can be approximated by an integral, and the normalised brane
distribution extracted by matching to supergravity:
\begin{equation}
\rho(u) = {{2}\over{m^2\gs}}\sqrt{a_0^2-u^2} \,,
\end{equation}
where the size of the \enh\ (in energy units) is
\begin{equation}
a_0=kL=mL^2=m\sqrt{{\gs N}\over{\pi}} \,.
\end{equation}
In the supergravity approximation, $\gs N$ is large, and so the size
of the \enh\ is much greater than the breaking parameter $m$.  

As emphasised in \cite{bpp}, this general method of analysis, matching
gravity and gauge theory $\tau$ functions, may provide clues to the
supergravity solution representing a more general gauge theory flow.
Work along these lines was undertaken in \cite{evansplus2}; some
subtleties remain to be understood.

\section{Outlook}

In both the \enh\ and the PW spacetimes, the supergravity description
turns out to be strongly coupled at the \enh\ locus; it is not clear
whether there is any weakly coupled description for the physics there.
The original \enh\ spacetime can also be studied without taking the
decoupling limit, and in that case the brane expansion mechanism of
singularity resolution can occur in a regime where the supergravity
description is everywhere weakly coupled.  Some progress was made in
\cite{jmpr} for finite temperature in that \net\ system, and more work
is still needed.

Going to finite temperature in systems exhibiting gravity/gauge
duality can be simpler than in the general case, because the gauge
theory may guide expectations on the gravity side.  Some progress has
been made in this way on gravity duals at finite temperature without
naked singularities, but only at high temperature
\cite{highT1}-\cite{highT5}.  It will be interesting to discover the
mechanisms for singularity resolution in those systems at low
temperature, and indeed for the more general cases.

To our knowledge, the only examples to date where singularity
resolutions are understood explicitly involve timelike and null
singularities.  The case of spacelike singularities is harder to
understand, and may be intertwined with the black hole information
problem.

In the Lorentzian AdS/CFT correspondence, the CFT at finite
temperature has been argued to correspond to an AdS-Schwarzschild
black hole.  Wick rotation in the bulk, something which does not
appear to make sense in general in quantum gravity, can in AdS/CFT be
defined via Wick rotation in the CFT \cite{bkl}.  It can then be
argued that the region behind the horizon of an AdS-Schwarzschild
black hole containing the spacelike singularity may not be represented
in the CFT, because that region is absent in the Euclidean
continuation.  On the other hand, in \cite{horoog} the opposite point
of view was taken.  There, for the (unexcited) AdS bulk it was pointed
out that the conformal isometry of the bulk spacetime dictates working
with global AdS and that this must include regions behind horizons.
The CFT then lives on S$^3$ not $\bR^3$.  This requirement changes the
global structure of the spacetime, which then cannot be obtained as
the near-horizon geometry of a bunch of D3-branes. Questions about
brane renditions of bulk locality and diffeomorphism invariance
remain.

Many more general questions may be asked.  One question is whether
maximal analytic extensions of supergravity geometries are physical.
It is sometimes argued that such extensions are necessary in order to
avoid conical singularities, but these are rather mild singularities
in string theory.  Another question is whether eternal black holes
exist.  (White holes probably do not exist because they are unstable
to collapsing on very short timescales.)  Another problem, related to
black hole complementarity, is to understand the physics of an
observer who has crossed the horizon and is falling toward a spacelike
singularity.  Cosmological singularities are more difficult yet as the
spacetimes are time-dependent.  Cosmological horizons also present
present basic difficulties \cite{raph,willy,lenny}.  According to
\cite{tomb} the initial ``big bang'' singularity has a quite different
character to a black hole interior even though their Carter-Penrose
diagrams may look similar.

A proposal has been made very recently \cite{juanavatar} that eternal
black holes in AdS may be understood via a direct product of two
entangled CFT's.  It will be interesting to understand whether this
can be generalised.

\section*{Acknowledgements}

Research of the author is supported by NSERC of Canada and the
University of Toronto.
The author thanks participants of the ITP conference ``Avatars of M
Theory'', 
Tom Banks, Raphael Bousso, Lenny Susskind and Arkady Tseytlin for
useful interactions.

\bibliographystyle{amsalpha}

\end{document}